\documentclass[reprint,amsmath,amssymb,pra,aps, nofootinbib,superscriptaddress]{revtex4-1}

\usepackage[T1]{fontenc}
\usepackage[utf8]{inputenc}
\usepackage{amsmath, amssymb}
\usepackage{bbold}
\usepackage{bm}
\usepackage{tipa}
\usepackage{blindtext}
\usepackage[dvipsnames]{xcolor}
\definecolor{dcyan}{RGB}{0,100,100}
\usepackage{graphicx}
\usepackage{tabularx}
\usepackage{xfrac}
\usepackage{units}
\usepackage{float}
\usepackage{mathtools}
\usepackage{xcolor}
\definecolor{green_cust}{RGB}{0,154,85}
\definecolor{red_cust}{RGB}{173,49,54}
\definecolor{blue_cust}{RGB}{0,103,148}
\usepackage{soul}
\usepackage{eqparbox}
\newcommand{\eqmathbox}[2][E]{\eqmakebox[#1]{$\displaystyle #2$}}

\newcommand{\ket}[1]{|{#1}\rangle}
\usepackage{hyperref}
\hypersetup{colorlinks=true, linkcolor=red_cust, citecolor=green_cust, filecolor=blue_cust, urlcolor=blue_cust}
\begin{document}
\title{Light-controlled strong coupling of optical cavity modes spaced by 200 THz}
\author{Lavanya Taneja}
\affiliation{The Department of Physics, The James Franck Institute, and The Pritzker School of Molecular Engineering, The University of Chicago, Chicago, IL}
\affiliation{The Department of Physics, Stanford University, Stanford, CA}
\author{David I. Schuster}
\affiliation{The Department of Applied Physics, Stanford University, Stanford, CA}
\author{Jonathan Simon}
\affiliation{The Department of Physics, Stanford University, Stanford, CA}
\affiliation{The Department of Applied Physics, Stanford University, Stanford, CA}
\date{\today}

\begin{abstract}
Cavities have driven significant advances in optical physics and quantum science, with applications ranging from lasers and spectroscopy to quantum information processing, simulation and metrology. For standard optical cavities, each eigenmode corresponds to a single, well-defined frequency. Here, we present a macroscopic optical Fabry-P{\'e}rot cavity whose eigenmodes are coherent superpositions of two frequency modes in the VIS-NIR range. Specifically, we demonstrate strong coupling between 384 THz (780~nm) and 580 THz (516~nm) cavity modes by incorporating an intracavity $\chi^{(2)}$ crystal driven by a non-resonant optical pump at 1529~nm. Strong coupling enables us to demonstrate frequency conversion with an end-to-end, free-space conversion efficiency of 30(1)\%, limited by current cavity design and internal cavity losses. We also demonstrate coupling between distinct spatial modes at the two frequencies, extending coherent control to the spatial basis. In conjunction with improved resonator design and low-loss nonlinear crystals, we anticipate a factor of $>$50 increase in two-mode-cooperativity for stronger coupling and near-unity conversion efficiency at low pump powers. This platform opens new avenues for cavity-QED experiments, with potential applications spanning cavity-mediated interactions between distinct atomic species, interconnects for quantum networking and modular computing, and spatially multimode cavity physics.
\end{abstract}

\maketitle
\section{Introduction}
\label{sec:intro}
Cavities have contributed to important advances in optics and quantum science, ranging from lasers and spectroscopy~\cite{adler2010CavityEnhanced,karpf2016Ultrasensitive, thorpe2008Cavityenhanced} to quantum information processing~\cite{welte2018PhotonMediated, law1997Deterministic, grinkemeyer2024ErrorDetected, yan2023Superradiant, liu2023Realization, hu2024Siteselective, deist2022MidCircuit}, simulation~\cite{periwal2021Programmable, cooper2024Graph, clark2020Observation}, and metrology~\cite{schleier-smith2011Cavityenabled, greve2022Entanglementenhanced, li2023Improving, panda2024Coherence}. Most typical applications of optical cavities rely on uncoupled, monochromatic cavity eigenmodes. Cavity eigenmodes at different frequencies, when coupled, can expand the potential of cavities, enabling applications such as generation of squeezed light~\cite{wade2015squeezed, pereira1988squeezed}, narrow bandwidth generation of entangled photon-pairs~\cite{ahlrichs2019TriplyResonant}, microwave to optical transduction~\cite{andrews2014Bidirectional, kumar2023Quantumenabled}, and optical frequency conversion using single-photon nonlinearities~\cite{wang2023Quantum, chen2021Photon, gao2021Broadband}. The frequency coupling is typically mediated by atomic ensembles~\cite{kumar2023Quantumenabled} or $\chi^{(2)}$ nonlinear materials~\cite{wade2015squeezed, pereira1988squeezed, 10.5555/1817101, ahlrichs2019TriplyResonant, andrews2014Bidirectional,wang2023Quantum, chen2021Photon, gao2021Broadband}. 

Strong, coherent coupling of optical cavity modes and consequent frequency conversion have previously been demonstrated in triply-resonant aluminum nitride~\cite{ramelow2019strong} and silicon nitride~\cite{guo1026onchip} microring resonators. The strong coupling condition is achieved when the coupling strength between two modes exceeds the loss rates of the two modes. The tight confinement of optical modes and longer interaction lengths in these nanophotonic platforms provide large coupling strengths at low pump powers~\cite{wang2023Quantum, chen2021Photon, gao2021Broadband}. However, compared to free-space cavities, these systems are limited by fiber-device coupling efficiencies~\cite{wang2023Quantum, chen2021Photon} and cannot accommodate quantum systems such as neutral atoms for cavity-QED experiments.

In this work, we present a macroscopic, doubly-resonant cavity with an intracavity nonlinear crystal to demonstrate strong coherent coupling between cavity modes at two frequencies. The coupling, mediated by a non-resonant, classical optical pump, provides hybridized Floquet eigenmodes at the two frequencies. As proof-of-concept, we demonstrate normal mode splitting of the cavity modes at 780~nm (384 THz) and 516~nm (580 THz). The coupling strength ($g$) is shown to exceed the dissipation rates ($\kappa$'s) in the system with a two-mode-cooperativity $\mathcal{C} = 4g^2/\kappa_1\kappa_2$ reaching 7. We utilize this mode coupling to demonstrate frequency conversion for a classical input with 30(1)\% end-to-end efficiency in free space, at $90$ mW of pump power. We also demonstrate coherent coupling between otherwise orthogonal spatial modes at the two frequencies. 

The paper is structured as follows: Section~\ref{sec:secB} describes the cavity and presents results demonstrating strong coherent coupling between two single-frequency cavity modes. Section~\ref{sec:secC} discusses the cooperative enhancement provided by the system and presents results on transduction and spatial mode coupling. Section~\ref{sec:secD} outlines routes to improve the performance of the platform toward higher cooperativities at lower pump powers, and discusses potential applications to atomic and classical optics platforms.

\begin{figure*} 
	\centering
	\includegraphics[width=0.9\textwidth]{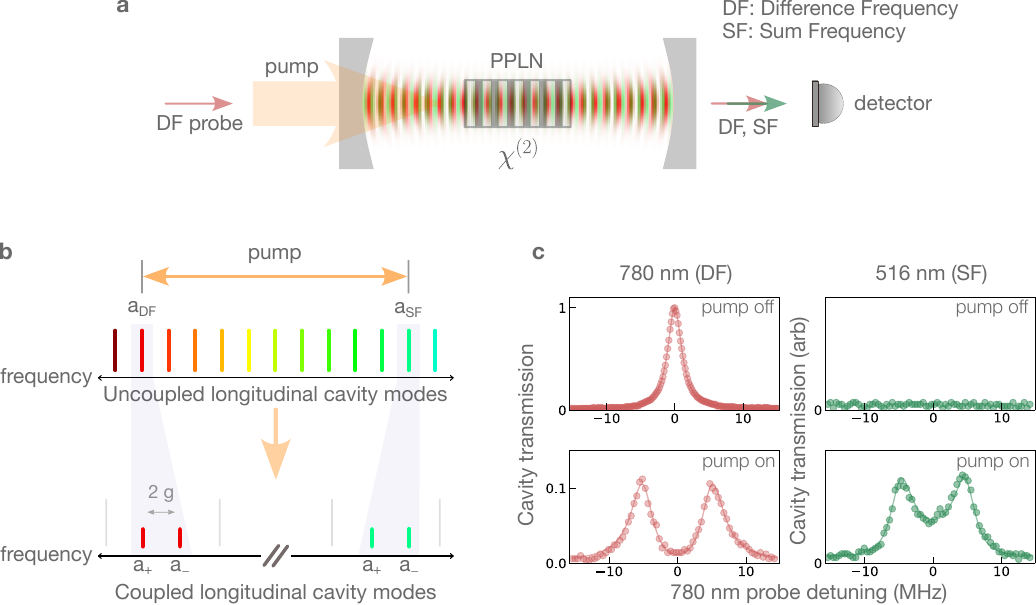}
	\caption{
    \textbf{Optical pump-mediated coupling of cavity modes at 780~nm and 516~nm}. \textbf{(a)} The experimental cavity: an AR-coated and 5\% MgO doped periodically-poled lithium niobate (ppLN) crystal is placed within a macroscopic Fabry-P{\'e}rot cavity which is doubly-resonant at 780~nm (difference-frequency or DF) and 516~nm (sum-frequency or SF) wavelengths. The 10 mm long crystal is positioned approximately at the center of the 82 mm long cavity (SI~\ref{SI:PhysicalSetup}) \textbf{(b)} When pumped at 1529~nm, the crystal couples the two cavities modes ($\hat{a}_{\text{DF}}$ and $\hat{a}_{\text{SF}}$) via sum- and difference-frequency generation processes. When the pump is near two-mode resonance, this coupling admixes the two modes yielding eigenmodes (with lowering operators $\hat{a}_+$ and $\hat{a}_-$) that are the superposition of the DF and SF modes. In the absence of loss, the minimum frequency splitting between the hybridized modes is determined by the coupling strength $g=\sqrt{\epsilon}~\nu$, where $\epsilon$ is the single-pass wavelength-conversion efficiency and $\nu$ is the cavity free-spectral range ($c/2L_{\text{cav}}$). \textbf{(c)} We measure the resonator transmission at DF and SF as a function of the detuning of the DF probe laser from the DF cavity resonance and observe a mode splitting in the spectrum when the crystal is driven by a pump. The frequency splitting is evidence of hybridization of the DF and SF modes. The solid lines are fits. The spectra correspond to a coupling $g = 2\pi \times 5.0(2)$ MHz and cavity decay rates $\{\kappa_{\text{DF}}, \kappa_{\text{SF}}\} = 2\pi \times \{2.5(1), 6(1)\}$ MHz, providing a cooperativity $\mathcal{C} = 4g^2/\kappa_{\text{DF}}\kappa_{\text{SF}} = 7(1)$. 
	}
	\label{fig:Fig1}
\end{figure*}

\section{Coupled cavity modes in a doubly-resonant cavity}
\label{sec:secB}
The experimental scheme is described in Fig~\ref{fig:Fig1}. We integrate a 5\% MgO-doped periodically-poled Lithium Niobate (ppLN) crystal within a macroscopic Fabry-P{\'e}rot optical cavity designed to be resonant at 780~nm and 516~nm. The crystal surfaces are AR coated at the two wavelengths to minimize cavity loss (SI~\ref{SI:PhysicalSetup}). The crystal is quasi-phase-matched for sum- and difference-frequency generation processes to couple 516~nm and 780~nm (henceforth, the sum-frequency or SF, and the difference-frequency or DF, respectively), using a pump at 1529~nm. 

In a frame rotating with the pump field, the Hamiltonian for the two cavity modes $\hat{a}_{\text{DF}}$ and $\hat{a}_{\text{SF}}$ coupled by the crystal is
\begin{equation}
   \begin{aligned}
    H&= -\delta_{\text{pump}} \hat{a}_{\text{SF}}^\dagger \hat{a}_{\text{SF}} + g (\hat{a}_{\text{DF}}^\dagger \hat{a}_{\text{SF}} + \hat{a}_{\text{DF}} \hat{a}_{\text{SF}}^\dagger)
\end{aligned}
\label{eq:hamiltonian}
\end{equation}

The coupling strength is $g = \sqrt{\epsilon}~\nu$ (\cite{steckQuantum}, SI ~\ref{SI:CouplingCalc}), where $\epsilon$ is the single-pass wavelength-conversion efficiency of the pumped crystal and $\nu = c/2L_{\text{cav}}$ is the cavity free-spectral range, where $L_{\text{cav}}$ is the length of the cavity. $\delta_{\text{pump}}$ is the detuning of the pump frequency from $\omega_p = \omega_{\text{SF}} -\omega_{\text{DF}}$, where $\omega_{\text{DF}}$  and $\omega_{\text{SF}}$ are the resonance frequencies of bare cavity modes $\hat{a}_{\text{DF}}$ and $\hat{a}_{\text{SF}}$. The single-pass conversion efficiency $\epsilon$ scales linearly with the pump power, leading to a square-root dependence of the coupling $g$ on the pump power.

The beamsplitter Hamiltonian in Eq.~\ref{eq:hamiltonian} describes two coupled bosonic modes with the pump frequency controlling the detuning of the drive between the DF and SF cavity modes, and the pump power controlling the coupling strength. Diagonalizing the Hamiltonian yields two eigenmodes that are superpositions of the uncoupled cavity eigenmodes of the form $\hat{a}_+ = (\cos\left(\frac{\theta}{2}\right)\hat{a}_{\text{DF}}+\sin\left(\frac{\theta}{2}\right) \hat{a}_{\text{SF}})$ and $\hat{a}_- = (-\sin\left(\frac{\theta}{2}\right)\hat{a}_{\text{DF}}+\cos\left(\frac{\theta}{2}\right) \hat{a}_{\text{SF}})$, where $\tan(\theta) = 2g/\delta_{\text{pump}}$. Thus, the new cavity quantizes the electromagnetic field in a way such that each eigenmode has two frequency components.
\begin{figure}
	\centering
	\includegraphics[width=0.42\textwidth]{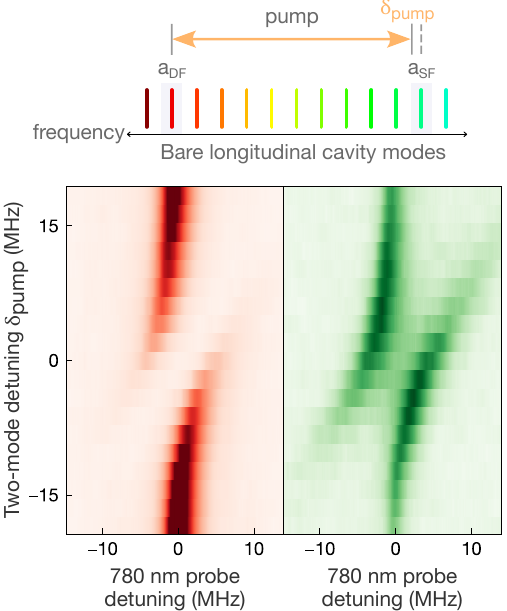}
	\caption{
		\textbf{Dependence on pump frequency.}  The contribution of the bare modes to the hybridized modes is controlled by the detuning $\delta_{\text{pump}}$ of the pump from the two-mode resonance. This is depicted in the cavity transmission at DF (left, red) and SF (right, green), when the 780~nm (DF) probe frequency and the two-mode detuning (set by the 1529~nm pump frequency) are simultaneously scanned. Away from two-mode resonance, the probe primarily couples to the weakly perturbed bare DF mode, resulting in lower transmission at SF. 
	}
	\label{fig:Fig2}
\end{figure}

At the two-mode resonance ($\delta_{\text{pump}}$ = 0), the coupling between the two frequency modes manifests spectroscopically as a frequency splitting of $2 g$ in the cavity transmission spectrum (Fig~\ref{fig:Fig1}(b)). Fig~\ref{fig:Fig1}(c) shows the observed spectrum at 780~nm and 516~nm when the cavity is probed using an input at 780~nm at a pump power of $\sim$200 mW. The symmetric splitting corresponds to $g = 2\pi\times 5.0(2)$ MHz, with total cavity linewidths of $\{\kappa_{\text{DF}}, \kappa_{\text{SF}}\} = 2\pi\times \{2.5(1), 6(1)\}$ MHz as inferred from the fits to the measured spectra (SI~\ref{SI:DrivenCavity}). 

The two-mode cooperativity $\mathcal{C} = \frac{4g^2}{\kappa_{\text{DF}} \kappa_{\text{SF}}}$ compares the coupling rate with the dissipation rates of the two modes. It provides information about the number of coherent oscillations that can take place for light between the DF and SF cavity modes before it is lost to the environment. A resolvable splitting in the transmission spectrum, corresponding to the condition $2g \geq(\kappa_\text{SF} + \kappa_\text{DF})/2$, implies a cooperativity $\mathcal{C}>1$, the strong-coupling condition of cavity QED. The frequency splitting in Fig~\ref{fig:Fig1}(c) corresponds to $\mathcal{C} = 7(1)$.

Figure~\ref{fig:Fig2} shows an avoided crossing in cavity transmission around the bare DF and SF modes as the cavity is probed with 780~nm light and the pump detuning $\delta_{\text{pump}}$ is scanned. The contributions of each of the bare modes toward the new eigenmodes vary with the pump detuning, resulting in the variation of the cavity transmission of the peaks. Away from the two-mode resonance condition, the probe couples to the very slightly perturbed bare DF mode, leading to strong DF transmission and weak SF transmission. The asymmetric transmission for the SF mode at a particular two-mode detuning arises due to different dissipation rates of the DF and SF modes (SI~\ref{SI:Asymmetry}).

For this section of the experiment, both end-mirrors have equal transmission coefficients and the cavity length is locked absolutely to a stable reference at 785~nm (SI~\ref{SI:PhysicalSetup}).

\begin{figure}
	\centering
	\includegraphics[width=0.42\textwidth]{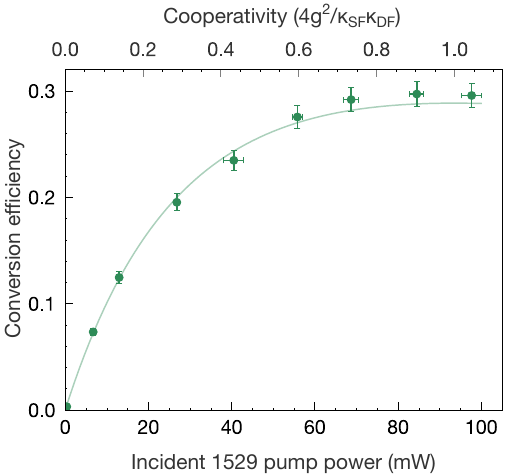}
	\caption{
    \textbf{780~nm $\rightarrow$ 516~nm end-to-end conversion efficiency}. The cooperative enhancement by the cavity at the input and output modes is utilized for optical transduction of a classical 780~nm input. The maximum end-to-end conversion efficiency is measured to be 30(1)\%, limited by cavity design and internal losses. The solid line is a one-parameter fit (SI~\ref{SI:conversioneff}) to the measured data. The maximum external efficiency for fixed external and internal losses is reached when the internal efficiency reaches unity, at $\mathcal{C} = 1$. The major source of error in the experiment were drifts of the input laser power. Note that a higher transmission cavity outcoupler (SI~\ref{SI:PhysicalSetup}) was used in this measurement for better impedance matching, with $\{\kappa_{\text{DF}}, \kappa_{\text{SF}}\} = 2\pi \times \{5.3(2), 12.3(6)\}$ MHz.}
	\label{fig:Fig3}
\end{figure}
\section{Optical transduction and spatial-mode coupling}
\label{sec:secC}
The cooperative enhancement from the two cavity modes can be used to obtain efficient, optically-tunable frequency conversion without the need for resonant enhancement~\cite{guo1026onchip} or waveguide confinement of the pump~\cite{bersin2024Telecom}. The optical pump modulates the cavity's optical path length at $c/\lambda_\text{pump} = 196$ THz, generating a frequency sideband at 516~nm. This modulation and the consequent frequency shifting can be Purcell-enhanced by tuning the input frequency and the output frequency (via the pump) into resonance with the cavity modes.

Using the input-output formalism for driven coupled cavity modes (SI~\ref{SI:DrivenCavity}), the conversion efficiency at resonance and zero probe detuning is
\begin{equation}
    \begin{aligned}
\eta = \Bigg\vert\frac{\hat{a}_{\text{2}}^{\text{out}}}{\hat{a}_{\text{1}}^{\text{in}}}\Bigg\vert^2  &= \frac{16g^2\kappa_{\text{DF}}^{\text{ext}}\kappa_{\text{SF}}^{\text{ext}}}{(4g^2 + \kappa_{\text{DF}}\kappa_{\text{SF}})^2} = \frac{4\mathcal{C}}{(1+\mathcal{C})^2}\frac{\kappa_{\text{DF}}^{\text{ext}}}{\kappa_{\text{DF}}}\frac{\kappa_{\text{SF}}^{\text{ext}}}{\kappa_{\text{SF}}}\\
\end{aligned}
\label{eq:conveff}
\end{equation}
where $\kappa^{\text{ext}}_{i} = T_i~\nu$ denote the external linewidths of the cavity modes, with $T_i =$ transmission coefficient of the coupling mirror at frequency $i$. Figure~\ref{fig:Fig3} shows the measured conversion efficiency at different pump powers and values of $\mathcal{C}$, with a maximum conversion efficiency of 30(1)\%. For fixed external and internal losses, the maximum efficiency is reached when $\mathcal{C} = 1$. In terms of cavity finesses $\mathcal{F} = \nu/\kappa$, $\mathcal{C}$ is equal to $4 \epsilon \mathcal{F}_{\text{DF}}\mathcal{F}_{\text{SF}}$, demonstrating that the `internal' conversion efficiency~\cite{wang2022Generalized} $4\mathcal{C}/(1+\mathcal{C})^2$ quickly approaches unity because of the $\mathcal{O}(10^4)$ enhancement of the single-pass conversion efficiency $\epsilon$ by the two cavity modes. The external conversion efficiency stops increasing after $\mathcal{C}$ reaches one because the photon extraction rate, dependent on outcoupler transmission, fails to keep up with the increasing internal coupling rate. The latter manifests as the splitting in the spectrum (Figs~\ref{fig:Fig1}(c) and~\ref{fig:Fig2}) because the occupation in the cavity modes begins to oscillate with frequency $2g$. For $\mathcal{C}>1$, the maximum conversion efficiency is reached away from resonance~\cite{wang2022Generalized}.

In the current experimental setup, the end-to-end conversion efficiency is limited (SI~\ref{SI:improv}) by the pump and cavity waists, the transmission of the out-coupling mirror, and ultimately, the intrinsic losses of the cavity (excluding in- and out-coupling) at the two wavelengths, quantified by $\kappa_{\text{DF}}^{\text{int}}$ and $\kappa_{\text{SF}}^{\text{int}}$. The optimum efficiency is achieved when the converter is ``impedance-matched'' and there is no reflection at the input (SI~\ref{SI:conversioneff},~\cite{wang2022Generalized}):
\begin{equation}
    \begin{aligned}
\frac{\kappa_{\text{DF}}^{\text{ext}}}{\kappa_{\text{DF}}^{\text{int}}} = \frac{\kappa_{\text{SF}}^{\text{ext}}}{\kappa_{\text{SF}}^{\text{int}}}  &=  \sqrt{1+\mathcal{C}_{\text{max}}},
\end{aligned}
\label{eq:optcoup}
\end{equation}
where $\mathcal{C}_{\text{max}} = 4g^2/\kappa_{\text{DF}}^{\text{int}}\kappa_{\text{SF}}^{\text{int}}$, is the cooperativity in the absence of any leakage through the mirrors. Given the constraints on $\kappa_{\text{DF}}^{\text{ext}}$ and $\kappa_{\text{SF}}^{\text{ext}}$, the optimum overall efficiency achievable is
$$
\begin{aligned}
\eta_\text{optimum} &\eqmathbox{=} \frac{2 + \mathcal{C}_{\text{max}} - 2 \sqrt{1 + \mathcal{C}_{\text{max}}}}{\mathcal{C}_{\text{max}}}\\
&\eqmathbox{\underset{\mathcal{C}_{\text{max}}\gg1}{\approx}}  1-\frac{2}{\sqrt{\mathcal{C}_{\text{max}}}}
\end{aligned}
$$
\begin{figure}
	\centering
	\includegraphics[width=0.42\textwidth]{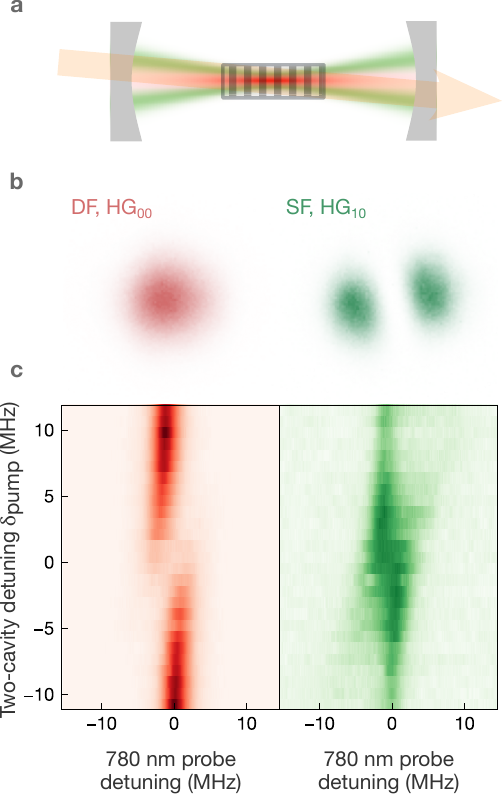}
	\caption{
		\textbf{Coupling orthogonal spatial modes across frequencies. (a)} The cavity length is stabilized to the DF HG$_{00}$ mode and the pump frequency is set at two-mode resonance with the SF HG$_{10}$ mode. The pump misalignment bridges the spatial orthogonality of the two modes. \textbf{(b)} Output of the cavity at DF (left, red) and SF (right, green) when probed at DF. \textbf{(c)} Avoided crossing in the transmission spectrum at DF (left, red) and SF (right, green) when probed with DF, corresponding to $g = 2\pi \times 2.7(1)$ MHz.}
	\label{fig:Fig4}
\end{figure}

The coupling $g$ between two frequency modes includes a dependence on the spatial overlap of the pump and cavity modes $\int_{V_{\text{crystal}}} E^{\ast}_\text{SF}(\vec{r}) E_\text{pump}(\vec{r}) E_\text{DF}(\vec{r}) dV$ over the crystal volume (SI~\ref{SI:CouplingCalc}). When the pump is not perfectly mode-matched due to suboptimal waists or misalignment, this can lead to overlap with, and hence, coupling to higher-order transverse modes of the cavity. However, the conversion to higher-order modes is Purcell suppressed via transverse mode non-degeneracy of the cavity, making the system more tolerant to small mismatches in mode-waists and misalignment of the pump. The reduction in coupling into the target mode can be offset via increased pump power.

In the presence of spatial overlap with higher order modes, the pump frequency can be varied to obtain cavity enhancement of coupling between different spatial modes at the sum- and difference- frequencies. As an example, Fig~\ref{fig:Fig4} demonstrates coherent coupling between a DF fundamental Gaussian (HG$_{\text{00}}$) mode and the spatially-orthogonal SF HG$_{\text{10}}$ mode, achieved by tilting the pump beam with respect to the cavity axis and varying the pump frequency to be resonant with the SF HG$_{\text{10}}$ mode. For higher coupling to targeted spatial modes, the pump can be beam shaped for more spatial overlap. The mode profile of the output is expected to be of high quality, as the output is spatially filtered by the cavity. 

\section{Outlook}
\label{sec:secD}
In this work, we have demonstrated optically controlled, strong coupling between two cavity modes spaced by 200~THz using an intracavity $\chi^{(2)}$ crystal, obtaining Floquet eigenmodes that exist at two frequencies simultaneously. Using the platform, we have presented a new approach to frequency conversion with low pump powers, reduced sensitivity to alignment and waist optimization, and negligible fiber-converter coupling losses due to near-perfect mode-matching of the input with the fundamental Gaussian cavity mode at DF. Extending coherent control to the spatial dimension, we have also demonstrated coupling between orthogonal spatial modes at the two frequencies by varying the spatial overlap of the input and output modes using the pump.

Looking ahead, we expect that improvements to the scheme such as utilizing lower-loss materials such as congruent ppLN~\cite{leidinger2015Comparative, schwesyg2011Optical} and potassium titanyl phosphate (KTP)~\cite{steinlechner2013Absorption} for cavity mode coupling, and redesigning the cavity for optimal waists, could increase the cooperativity $\mathcal{C}_{\text{max}}$ by a factor of $>$50 (SI \ref{SI:improv}) at current pump powers, enabling faster coupling between modes, higher cooperativity and higher conversion efficiencies at lower pump powers. We anticipate $\sim$70\% conversion efficiency at $\sim$60~mW of pump power with optimized waists, and $>90$\% conversion efficiency at lower pump powers using materials with lower losses (SI \ref{SI:improv}). 

More broadly, this work opens the door to cavity QED experiments with multiple atomic species~\cite{anand2024dualspecies, ramette2022AnyToAny} for gate and memory architectures~\cite{thompson2006HighBrightness, reim2011SinglePhotonLevel}, and developing quantum interconnects for atomic and hybrid platforms~\cite{ritter2012elementary, RevModPhys.87.1379, li2024Highrate}. In conjunction with experiments deploying cavities with small waists for readout of atoms and atom arrays~\cite{shadmany2024Cavity, chen2022High, li2024Highrate}, this platform could enable fast, optically-switchable readout to telecom wavelengths. The spatial control across frequencies could be useful for simultaneous trapping of ground state and Rydberg state atoms~\cite{kuga1997Novel, saffman2010Quantum}, and increased channel capacity for communication through few-mode optical fibers~\cite{zhu2018Orbital} and free-space~\cite{willner2017Recent, huang2014100, zhu2018Orbital, zhou2016Orbital, steinlechner2016Frequency}. Beam-shaping of the pump can also enable fast, optically-generated local potentials for cavity photons~\cite{klaers2010Bose}. Finally, the coupled mode cavity could enable new protocols for cavity state preparation and transfer~\cite{castanos-cervantes2024Coherent,miladinovic2011Adiabatic, larson2005cavity}.

\section{Acknowledgments}
This work was supported by AFOSR grant FA9550-22-1-0279, ARO grant W911NF-23-1-0053, AFOSR MURI grant FA9550-19-1-0399, and NSF QLCI-HQAN grant 2016136. We thank Tingran Wang and Zeyang Li for assistance with instrumentation, Ash Kumar for discussion of theory, and Zeyang Li and Adam Shaw for feedback on the manuscript. 

\section{Author Contributions}
The experiments were designed by L.T., J.S. and D.S. The construction of the apparatus and data collection were handled by L.T. All authors analyzed the data and contributed to the manuscript.
\vspace{0.15cm}
\section{Competing Interests}
The authors declare no competing financial or non-financial interests.

\section{Data Availability}
The experimental data presented in this manuscript are available from the corresponding author upon request, due to the proprietary file formats employed in the data collection process.
\section{Code Availability}
The source code for simulations throughout are available from the corresponding author upon request. 
\section{Additional Information}
Correspondence and requests for materials should be addressed to J.S. (jonsimon@stanford.edu). Supplementary information is available for this paper.
\bibliographystyle{naturemag}
\bibliography{references0}
\onecolumngrid
\clearpage
\newpage
\onecolumngrid
\newpage
\section*{Supplementary Information}
\appendix
\renewcommand{\appendixname}{Supplement}
\renewcommand{\theequation}{S\arabic{equation}}
\renewcommand{\figurename}{Supplemental Information Fig}
\renewcommand{\tablename}{Table}
\setcounter{figure}{0}
\renewcommand{\thefigure}{S\arabic{figure}}
\newcounter{SIfig}
\renewcommand{\theSIfig}{S\arabic{SIfig}}
\setcounter{table}{0}

\section{Construction of the cavity}
\label{SI:PhysicalSetup}

The setup uses a 2~mm~x~1~mm~x~10~mm, 5\% MgO doped lithium niobate crystal (HC Photonics) periodically poled for SFG: 1529~nm (pump) and 780~nm $\rightarrow$ 516~nm and DFG: 1529~nm (pump) and 516~nm $\rightarrow$ 780~nm. The surfaces of the crystal are AR coated (FiveNine Optics) for 780~nm (R $\sim 0.0005$) and 516~nm (R $\sim 0.0004$). The crystal surface reflects $\sim30\%$ of pump light (1529~nm) per surface.

\begin{figure}[htbp]
	\includegraphics[width=\textwidth]{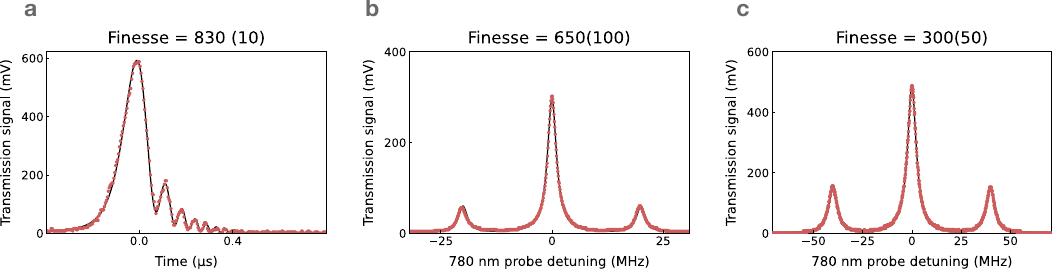}
	\caption{
		\textbf{Measurement of cavity finesse with intracavity 5\%-MgO doped ppLN crystal at 780~nm. (a)} Measurement via a cavity ringdown measurement~\cite{poirson1997Analytical}. The cavity transmission, when the probe frequency sweep rate $>\kappa/\mathcal{F}$, displays an oscillating behavior which can be fit to an exp-erfc function. The measured finesse $\mathcal{F} = 830(10)$ suggests a round-trip loss of 0.76\% in the crystal cavity. The error is calculated from the fit. \textbf{(b), (c)} Finesse measurements by modulating the probe using an EOM for (b) avoided-crossing measurements, and (c) conversion efficiency measurements with higher transmission outcoupler. The errors are calculated from uncertainties in cavity length measurement and x-axis calibration. The difference in (a) and (b) is most likely due to contamination in the course of the experiment.
	}
	  \refstepcounter{SIfig}\label{fig:SIFig1}
\end{figure}

The crystal is temperature stabilized at $\approx 35^{\circ}$ C. It is mounted on a 4-axis translation stage between two curved mirrors (ROC 100 mm) that form a standing-wave cavity at 780~nm and 516~nm. The cavity is 82 mm long, and its finesse at 780~nm is measured to be $\approx 830$ via cavity-ringdown measurements (Fig~\ref{fig:SIFig1}a), with a crystal limiting finesse of $\sim$900. The linewidth presented in the main text suggests a slightly larger loss (Fig~\ref{fig:SIFig1}b), potentially due to contamination in the course of the experiment. The cavity waists are designed for $110~\mu$m at 780~nm and $90~\mu$m at 516~nm, and the pump waist is estimated to be $\sim 116~\mu$m at the crystal. The 1529~nm pump is sourced from an EDFA. Dichroics are used to combine the pump and the DF, and to separate the pump, DF and SF for detection. 

\textbf{Avoided-crossing experiments:} The mirrors are coated at R = 99.965\% for 780~nm and 99.963\% for 516~nm leading to an undercoupled resonator setup. The mirrors are coated for high transmission at 1529~nm (R $= 0.02\%$). 

The cavity length is stabilized to a tracer 785~nm laser, which has been co-locked with the probe carrier using a separate stable transfer cavity. We sweep the probe frequency using an electro-optic modulator driven by a crystal oscillator and vary the detuning of the two cavities via current (hence, frequency) sweep of the 1529~nm pump. 

\textbf{Frequency conversion experiment:} To obtain higher end-to-end conversion efficiency, the mirror used as the 780~nm incoupler and 516~nm outcoupler is coated for 1.34\% transmission at 780~nm and 2.36\% transmission at 516~nm. The other end-mirror is the same from the avoided-crossing experiment, functioning as the blocked end of the cavity. The small leakage at 780~nm from this highly reflective mirror is used to stabilize the cavity to 780~nm resonance. The outcoupler is unintentionally coated for 98.6\% reflection at 1529~nm, which at perfect alignment should increase $g$.  We did not characterize the alignment of our pump beam with respect to the crystal, but did observe a greater $g$ for this section of the experiment for a given pump power ($\sim2\pi \times 5$ MHz at $\sim100$ mW compared to $\sim2\pi \times 5$ MHz at $\sim 200$ mW for the avoided-crossing experiment). The powers at 780~nm and 516~nm are measured t the input and output of the cavity using a NIST/PTB calibrated Thorlabs S120C sensor after bandpass filters at the respective wavelengths. The ratio is converted to a photon flux ratio by multiplying by $\lambda_{\text{SF}}/\lambda_{\text{DF}} = 516.5/780.2$. 

\section{Phase matching}
\label{SI:phasematching}
\begin{figure}[htbp]
	\includegraphics[width=\textwidth]{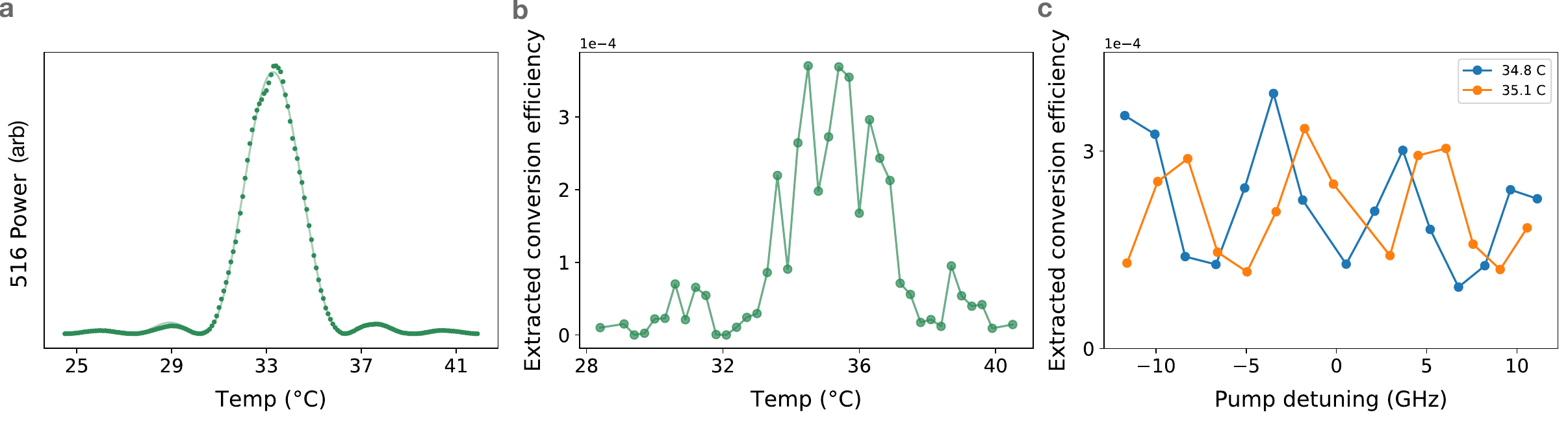}
	\caption{
		\textbf{Phase matching characterizations} \textbf{(a)} Out of cavity, we send 780~nm light into the crystal and measure the converted 516~nm power to characterize the crystal temperature dependence of frequency conversion. The data is fit to a sinc-squared function. \textbf{(b)} We place the crystal within the cavity and extract the single-pass conversion efficiency from the splitting in the spectrum ($g = \sqrt{\epsilon}~\nu$). In this measurement, the temperature is scanned and the pump frequency is kept fixed. The difference in optimal temperature is likely due to different crystal angles. The anomalous dips appear due to a crystal etalon effect for the pump (see text), confirmed by a pump frequency scan in \textbf{(c)} at fixed temperature.
	}
	\refstepcounter{SIfig}\label{fig:SIFig2}
\end{figure}

The periodic poling of the crystal is designed to provide quasi-phase matching at $35^{\circ}$ C. Fig~\ref{fig:SIFig2}(a) shows the crystal temperature dependence of converted power in free-space. The translation of the peak from designed temperature is likely due to angle of incidence deviation from $0^{\circ}$. Once the crystal is placed inside the cavity, we observed a maximum around $34.9^{\circ}$C (Fig~\ref{fig:SIFig2}(b)). Since our crystal surfaces are $\sim$30\% reflective at 1529~nm, the pump wavelength, there is an etalon effect leading to periodic variations in intracavity pump power, which leads to dips in Fig~\ref{fig:SIFig2}(b). With the temperature fixed, when the pump frequency is scanned (Fig~\ref{fig:SIFig2}(c)), the frequency spacing of the oscillations is $\sim4\times$ cavity FSRs = 6.4 GHz, which matches the crystal etalon free-spectral range after accounting for the refractive index of lithium niobate.

\section{Calculation of coupling \texorpdfstring{$g$}{g}}
\label{SI:CouplingCalc}

The expression for cavity coupling $g=\sqrt{\epsilon}~\nu$ can be derived using a classical field coupled-cavity calculation~\cite{steckQuantum}, where the fields are eventually replaced by quantized modes of the cavity. The expression generally holds true for any two cavity modes coupled by a coupler of efficiency $\epsilon$ and has been experimentally demonstrated for near-degenerate coupled resonators~\cite{stone2021Optical, miller2015Tunable}. 

Here, we outline the calculation of $g$ from first principles using the Hamiltonian for a far-detuned system of three-level atoms (states $\ket{G}, \ket{E}, \ket{R}$) with two cavity modes $a$ and $b$ in the rotating frame. The transitions $\ket{G}\rightarrow\ket{E}$, $\ket{E}\rightarrow\ket{R}$ and $\ket{R}\rightarrow\ket{G}$ correspond to $a$, pump and $b$ respectively. 

\begin{align*}
    H &= H_o + H_{int}\\
    H_o &= \delta_a a^\dagger a + \delta_b b^\dagger b + \delta_E E^\dagger E + \delta_R R^\dagger R \\
    H_{int} &= g_{a} a E^\dagger + \Omega_p R^\dagger E + g_b b^\dagger R + h.c. 
\end{align*}

where, $\delta_E$ is the single-photon detuning, $\delta_R$ is the two-photon detuning, $g_{a,b}$ are the atom-cavity coupling terms, $\Omega_p$ is the Rabi-frequency corresponding to the pump. Operator evolution equations after introducing loss and input:

    \begin{align}
    \dot{a} &= (i \delta_a - \frac{\kappa_a}{2}) a  - i g_{a} E + \sqrt{\kappa_a^{\text{ext}}} a_{in} \\
    \dot{E} &= (i \delta_E - \frac{\Gamma_E}{2}) E - i g_{a} a - i \Omega_p R \label{eq:Edot}\\
    \dot{R} &= (i \delta_R - \frac{\Gamma_R}{2}) R - i \Omega_p E - i g_b b  \label{eq:Rdot}\\
    \dot{b} &= (i \delta_b - \frac{\kappa_b}{2}) b - i g_b R + \sqrt{\kappa_b^{\text{ext}}} b_{in}
    \end{align}
where $\kappa$'s and $\Gamma$'s are the relevant losses. Since this is a far-detuned nonlinear process, we can adiabatically eliminate the upper states of the mediator by setting equations~\ref{eq:Edot} and~\ref{eq:Rdot} to 0. Using the expressions for $E$ and $R$ obtained from this step, we get an effective coupling between $a$ and $b$ reducing the Hamiltonian to
\begin{align*}
    H_{int} &= g (a^\dagger b + a b^\dagger)\\
    \text{where } g &= \frac{g_a g_b \Omega_p}{\delta_E \delta_R} 
\end{align*}
It is assumed that $\delta_E, \delta_R \gg \Omega_p, \Gamma_R, \Gamma_E$. For an atom at $\vec{r}_i$ in the crystal,
\begin{align*}
    g_\alpha (\vec{r}_i) &= \frac{\mu_\alpha}{\hbar}\sqrt{\frac{\hbar \omega_\alpha}{2 \epsilon_\alpha A_\alpha  L}} f_\alpha(\vec{r}_i)\\
    \Omega_p (\vec{r}_i)&= \frac{\mu_p E_{pump}(\vec{r}_i)}{\hbar}
\end{align*}
where $A_\alpha$ is the effective mode area for a travelling wave mode, $L$ is the length of the cavity, and $f_\alpha(r)$ incorporates the spatial variation of the mode. Integrating over all the atoms in the crystal volume,
\begin{align*}
    g = \left(2\chi^{(2)} \frac{\sqrt{\omega_a\omega_b}}{c\sqrt{n_an_bn_c}} L_{\text{crystal}}\sqrt{\frac{P_{\text{pump}}}{\epsilon_0 c}}\right) \left(\frac{1}{\sqrt{n_an_b}L_{\text{crystal}}\sqrt{A_a A_b (\pi w_p^2)}}\int_{V_{\text{crystal}}} dV f_a(\vec{r})f_b(\vec{r}) f_{pump} (\vec{r})\right) \frac{c}{2 L}
\end{align*}
where the integral includes the phase matching factor. Identifying the expression for the second-order susceptibility~\cite{10.5555/1817101} and the free-spectral ranges leaves us with
\begin{equation}
    \begin{aligned}
         g &= \sqrt{\epsilon'}~\Theta~  \frac{c}{2 L}=\sqrt{\epsilon}~\nu\\
    \end{aligned}
    \label{eq:coupling}
\end{equation}

where $\epsilon$ can be interpreted as the single-pass conversion efficiency of the crystal $\epsilon'$~\cite{10.5555/1817101} modified by the spatial overlap factor $\Theta$, and $\nu$ is the cavity free-spectral range. Note: (1) This model assumes that while the conversion takes place only in the crystal volume, the entire cavity mode exists in the dielectric medium with refractive indices equal to that of the crystal. The mode amplitude should be adjusted to account for the part-dielectric cavity. It, however, captures the dependence of coupling on beam waists which affects the spatial overlap (Figure~\ref{fig:SIFig3}a and SI~\ref{SI:improv}). (2) For a more accurate modeling of the crystal $\chi^{(2)}$, other permutations of atomic levels need to be included~\cite{10.5555/1817101}. This does not affect the final expression for $g$.

\section{Driven cavity equations for coupled cavity modes}
\label{SI:DrivenCavity}
From the simplified Hamiltonian of SI~\ref{SI:CouplingCalc}, the Langevin equations for coupled cavity modes $a$ and $b$ (single-ended cavity) are written as~\cite{steckQuantum}
$$
\begin{aligned}
\partial_t a(t) &= -i \omega_{0, a} a(t) - \frac{\kappa_a}{2}a(t) -i \mathcal{E} e^{-i\omega_L t} -\sqrt{\kappa_a^{\text{ext}}} a_{in}(t)-i g e^{i \omega_{p}t} b(t)\\
\partial_t b(t) &= -i \omega_{0, b} b(t) - \frac{\kappa_b}{2}b(t)-\sqrt{\kappa_b^{\text{ext}}} b_{in}(t)-i g e^{-i \omega_{p}t}a(t)
\end{aligned}
$$
where $\mathcal{E} e^{-i\omega_L t}$ is the coherent drive term with frequency $\omega_L$ for cavity mode $a$, \{$\omega_{0,i}$, $\kappa_i$ and $\kappa_i^{\text{ext}}$\} are the resonance frequency, total linewidth and external linewidth for mode $i$. $ge^{i\omega_p t}$ is the coupling term introduced by the pump. We define rotating frame operators:
$$
\begin{aligned}
\tilde a &\coloneqq a e^{i\omega_L t}\\
\tilde b &\coloneqq b e^{i(\omega_L+\omega_{p}) t}
\end{aligned}
$$
With the external bath in vacuum state (hence $\langle a_{in} \rangle, \langle b_{in} \rangle = 0$), at steady state:
$$
\begin{aligned}
i (\omega_L - \omega_{0, a}) \tilde a -i \mathcal{E} - \frac{\kappa_a}{2}\tilde a -i g  \tilde b &= 0\\
i (\omega_L + \omega_{p} - \omega_{0, b}) \tilde b  - \frac{\kappa_b}{2}\tilde b-i g \tilde a &= 0
\end{aligned}
$$
Let $\omega_L - \omega_{0,a} = \delta_{L,a0}$ and $\omega_{p} = \omega_{0,p}+\delta_{pump}$. With $\omega_{0,a}+\omega_{0,p}=\omega_{0,b}$,
$$
\begin{aligned}
i \delta_{L,a0} \tilde a -i \mathcal{E} - \frac{\kappa_a}{2}\tilde a -i g  \tilde b &= 0\\
i (\delta'_{L,a0}+\delta_{\text{pump}}) \tilde b  - \frac{\kappa_b}{2}\tilde b-i g \tilde a &= 0
\end{aligned}
$$
The input-output relations provide the output leaking out of the cavity for the two modes:
\begin{equation}
    \begin{aligned}
\langle a_{out} \rangle &= \mathcal{E}e^{-i\omega_L t}/\sqrt{\kappa_a^{\text{ext}}} + \sqrt{\kappa_a^{\text{ext}}} \langle \tilde a\rangle e^{-i\omega_L t}\\
\langle b_{out}\rangle &=  \sqrt{\kappa_b^{\text{ext}}} \langle \tilde  b\rangle e^{-i(\omega_L+\omega_p) t}\\
    \end{aligned}
    \label{eq:inputoutput}
\end{equation}
These relations provide expressions that are used to fit data shown in main text figures~\ref{fig:Fig1}(c)-\ref{fig:Fig4}. The $\kappa_{\text{DF}}$ is determined independently from the data and fits shown in Fig~\ref{fig:SIFig1}(b) and~\ref{fig:SIFig1}(c).

\section{Interpreting the avoided-crossing diagram for 516~nm}
\label{SI:Asymmetry}
Each horizontal slice on the avoided-crossing diagram in the main text figure~\ref{fig:Fig2} is the output of the cavity when it is probed with 780~nm light. At the two-mode resonance, we measure two peaks with equal transmission. Away from resonance, we measure more 780~nm transmission corresponding to the mode that has more contribution from the uncoupled DF eigenmode. The near-Lorentzian measurement of the off-resonant peaks is truncated in transmission due to the vertical resolution of the oscilloscope. 

Simultaneously, we also measure output at 516~nm, the SF wavelength (main text figure~\ref{fig:Fig2}, right column). If the dissipation of the two modes are equal, we expect equal transmissions for the split peaks in the SF transmission. This is because even with a lower SF contribution in an eigenmode, the excitation amplitude from the DF probe is stronger due to larger DF mode contribution. The two peak transmissions in the 516~nm spectrum should simultaneously decrease as the pump is tuned away from the two-mode resonance, because the eigenmodes that are excited more have less and less contribution from the SF mode. In this case, however, we see asymmetric transmission for the two 516~nm peaks at each horizontal slice away from resonance. This is due to a higher dissipation rate of the SF cavity mode than of the DF cavity mode, leading to seeing more loss (hence lower SF transmission) for coupled modes that have higher contribution of the bare SF mode.

\section{Coupled cavity conversion efficiency and impedance matching}
\label{SI:conversioneff}
The data shown in main text figure~\ref{fig:Fig3} is fit to equation~\ref{eq:conveff}, using $\kappa^{\text{ext}}_i$ values from mirror coating specifications and $\kappa_i$ values from avoided-crossing data for the cavity with a higher transmission outcoupler. A single fit parameter was used to account for the frequency-dependent variation in single-pass conversion efficiency at equal pump powers (as discussed in SI~\ref{SI:phasematching}, figure~\ref{fig:SIFig2}(c)).

For fixed internal losses and coupling $g$, the highest conversion efficiencies are achieved when there is zero reflection from the input port. The expression for the reflection is
$$
\begin{aligned}
\Bigg\vert\frac{\hat{a}_{\text{DF}}^{\text{out}}}{\hat{a}_{\text{DF}}^{\text{in}}}\Bigg\vert^2  &= \Big(\frac{\mathcal{C}_\text{max}+(\beta_{\text{DF}}-1)(1+\beta_{\text{SF}})}{\mathcal{C}_\text{max}+(1+\beta_{\text{DF}})(1+\beta_{\text{SF}})}\Big )^2\\
\end{aligned}
$$
where $\beta_i = \frac{\kappa_{i}^{\text{ext}}}{\kappa_i^{\text{int}}}$. 

Physically, the condition $\vert \hat{a}_{\text{DF}}^{\text{out}} \vert^2 = 0$ requires the input rate of photons ($\kappa_{\text{DF}}^{\text{ext}}$) should be equal to the sum of the loss rate from the first cavity ($\kappa_{\text{DF}}^{\text{int}}$) and the rate of coupling to the other cavity ($\frac{4g^2}{\kappa_{\text{SF}}}$) 
$$
\begin{aligned}
\kappa_{\text{DF}}^{\text{ext}} &= \kappa_{\text{DF}}^{\text{int}} + \frac{4g^2}{\kappa_{\text{SF}}}\\
(\kappa_{\text{DF}}^{\text{ext}}-\kappa_{\text{DF}}^{\text{int}}) (\kappa_{\text{SF}}^{\text{ext}}+\kappa_{\text{SF}}^{\text{int}})&= 4g^2\\
\end{aligned}
$$
Since the converter should be symmetric in operation, we can interchange $a$ and $b$ to write
$$
\begin{aligned}
(\kappa_{\text{SF}}^{\text{ext}}-\kappa_{\text{SF}}^{\text{int}}) (\kappa_{\text{DF}}^{\text{ext}}+\kappa_{\text{DF}}^{\text{int}})&= 4g^2\\
\end{aligned}
$$
which implies
$$
\begin{aligned}
\frac{\kappa_{\text{DF}}^{\text{ext}}}{\kappa_{\text{DF}}^{\text{int}}} = \frac{\kappa_{\text{SF}}^{\text{ext}}}{\kappa_{\text{SF}}^{\text{int}}} &= \sqrt{1+\frac{4g^2}{\kappa_{\text{DF}}^{\text{int}}\kappa_{\text{SF}}^{\text{int}}}}, \mathrm{or}\\
\beta_{\text{DF}} = \beta_{\text{SF}} &= \sqrt{1+\mathcal{C}_\text{max}}
\end{aligned}
$$

\section{Estimated improvements}
\label{SI:improv}
\begin{figure}[htbp]
	\includegraphics[width=\textwidth]{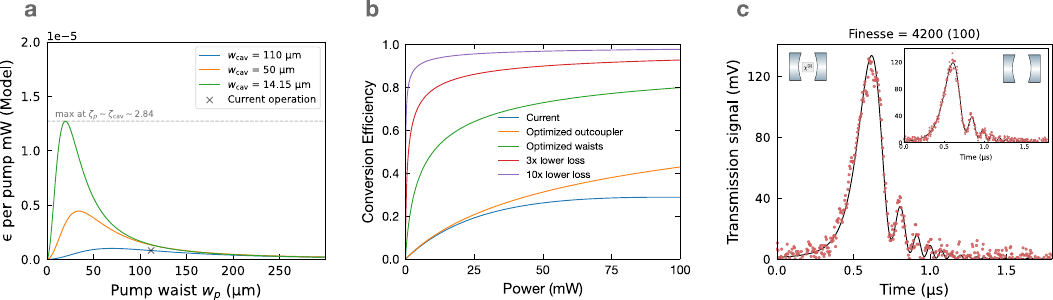}
	\caption{
		\textbf{Estimated improvements (a)} Calculation for the single-pass efficiency per mW of pump power for the pump and input beam sizes for the current cavity design (with 780~nm cavity waist $\sim$110 $\mu$m at DF), intermediate waist ($\sim$50 $\mu$m) and optimal waist ($\sim$14 $\mu$m). The maximum is obtained when the parameters $\zeta$ of the cavity and the pump are equal to 2.84, where $\zeta = L_{\text{crystal}}/2z_R$ and $z_R$ is the Rayleigh length of the pump and the cavity. The calculations are done for 1 mW of pump power using the expression in SI~\ref{SI:CouplingCalc}. \textbf{(b)} Expected end-to-end conversion efficiency in different scenarios. Blue: fit corresponding to data in main text Fig~\ref{fig:Fig3}. Orange: with optimized outcouplers with the current crystal and cavity/pump waists. Green: with optimized waists and outcouplers with the current crystal. Red: with optimized waists and outcouplers and 3x lower losses at both SF and DF. Purple: with 10x lower losses at both SF and DF. \textbf{(c)} Finesse measurements with an intracavity undoped, unpoled 3~mm long lithium niobate crystal provided a limiting cavity finesse of 9900 calculated from cavity finesse without crystal (7400) and cavity finesse with crystal (4200).
	}
	\refstepcounter{SIfig}\label{fig:SIFig3}
\end{figure}

A gain in cooperativity is expected from optimizing pump- and cavity waists at the crystal to increase the single-pass efficiency. From the model, we see that the optimal point of operation is when $L_{\text{crystal}}/2 z_R = 2.84$, where $z_R$ is the Rayleigh range of the cavity mode/pump beam. This is consistent with the theoretical extension of the Boyd-Kleinman theory for non-resonant SFG by Guha et al~\cite{guha1980effects}, where the maximum spatial overlap between the pump and the input, and the maximum single pass efficiency is ensured at $L_{\text{crystal}}/2 z_R = 2.84$ in the absence of walk-off. Optimizing the pump waist in our current setup to $20~\mu$m and designing a running-wave cavity~\cite{chen2022High, jaffe2021Aberrated, li2024Highrate} with a waist of $\sim14~\mu$m is expected to provide an enhancement of a factor of $\sim15$ in the single-pass conversion efficiency (Figs~\ref{fig:SIFig3}(a) and~\ref{fig:SIFig3}(b)). 

For frequency-conversion with the current internal cavity losses, increasing the outcoupler transmission and pump power can provide higher efficiencies (Fig~\ref{fig:SIFig3}(b)). In addition, exploring lower-loss nonlinear materials could be useful. Experiments with intracavity undoped lithium niobate have provided limiting cavity finesse of 9900 (calculated from finesse measurement with and without the crystal) (Fig~\ref{fig:SIFig3}(c)). The crystal length was 3 mm (Manufacturer: Castech), suggesting an absorption loss of 0.001 per cm. Measurements of congruently grown lithium niobate suggest losses as low as 0.0004 per cm~\cite{leidinger2015Comparative} at 780~nm, suggesting a 10x gain in cooperativity from lower losses at 780~nm. Moving to telecom wavelengths from 516~nm would provide a further gain in cooperativity due to lower losses in lithium niobate in the telecom frequency range. One possible combination is 1570~nm as the pump, and 1550~nm and 780~nm as the difference- and sum- frequencies.

Other crystals, such as potassium titanyl phosphate (KTP), exhibit lower absorption losses than lithium niobate (LN). However, due to their lower electro-optic coefficients (around half of LN for KTP), they require a longer interaction length. After accounting for the length, ppKTP should provide a factor of $\sim 7$ improvement in loss over 5\%-MgO doped LN at 780~nm, but evaluates to comparable overall losses as compared to congruent LN (Loss of ppKTP at 775 nm has been measured to be $\sim$127 ppm per cm~\cite{steinlechner2013Absorption}, $\sim$3435 ppm per cm for 5\%-MgO doped LN~\cite{schwesyg2011Optical} and 400 ppm per cm for congruently grown LN~\cite{leidinger2015Comparative}).  

\twocolumngrid

\clearpage

\end{document}